\documentstyle[]{mn}

\title[TAURUS observations of NGC~4151]{Kinematics of Ionised Gas in the Barred Seyfert Galaxy NGC~4151}

\author[M. W. Asif et al.]
        {M. W. Asif$^{1}$, C. G. Mundell$^{2}$ \& A. Pedlar$^{1}$\\
        \\
$^{1}$Jodrell Bank Observatory, School of Physics and Astronomy, 
University of Manchester, Macclesfield,
Cheshire SK11~9DL\\
$^{2}$Astrophysics Research Institute, Liverpool John Moores University, Twelve Quays
House, Birkenhead L41~1LD\\}

\begin{document}
\maketitle

\begin{abstract}
\large

{We have determined the structure and kinematics of ionised gas in the 
weak oval bar of the archetypal Seyfert 1 galaxy, NGC~4151, using the 
 TAURUS Fabry-Perot interferometer to simultaneously map the
 distribution and kinematics of H$\beta$ emission. We also present
 broad-band ultraviolet imaging of the host galaxy, obtained with
 XMM-Newton, that shows the detailed distribution of star formation in
 the bar and in the optically-faint outer spiral arms. We compare
 the distribution and kinematics of ionised gas with that previously
 determined in neutral hydrogen by Mundell \& Shone; we suggest that the
 distribution of bright, patchy UV emission close to the H{\sc i}
 shocks is consistent with ionisation by star clusters that
 have formed in compressed pre-shock gas. These clusters then travel
 ballistically through the gaseous shock to ionise gas downstream
 along the leading edge of the
 bar. In addition, we detect, for the first time, ionised gas within the shock itself
 which is streaming to smaller radii in the same manner as the neutral gas.}
\normalsize

\end{abstract}

\section{Introduction}

          
Present in at least 58\% of nearby galaxies, bars are thought to be
ubiquitous in galaxies in the local Universe (e.g. Mulchaey \& Regan
1999; Menendez-Delmestre et al. 2004). Bars may play a key role in triggering
star formation (e.g. Martinet \& Friedli 1997; Reynaud \& Downes 1999;
Sheth et al. 2002) and at higher redshift, may form bulges and drive
galaxy evolution (e.g. Norman, Sellwood \& Hasan 1996). Their role in
the triggering and fuelling of nuclear activity in Active Galactic
Nuclei (AGN) is still unclear but bars may offer a useful mechanism
for transporting gas towards the centres of galaxies (Roberts, Huntley
\& Van Albada 1979; Schwarz 1985; Shlosman, Frank \& Begelman 1989;
Sakamoto et al. 1997), for example when gas streamlines associated
with different orbit families converge and form shocks that dissipate
angular momentum, thereby allowing gas to move to smaller radii
(Prendergast 1983; Athanassoula 1992a,b).  Dust lanes were
long-considered circumstantial evidence for the presence of such
gaseous shocks but direct kinematic evidence was difficult to obtain
due to the small amounts of ionised gas associated with them
(e.g. Lindblad \& Joers\"ater 1988; Lindblad, Lindblad \& Athanassoula
1996). Subsequent high angular resolution radio studies of neutral
hydrogen in the nearby barred Seyfert, NGC~4151
 (Mundell \& Shone 1999) provided strong kinematic evidence for
 streaming shocks in a weak barred potential, as predicted by
 numerical simulations (Athanassoula 1992a,b).

NGC~4151 is a nearby grand-design barred spiral galaxy, classified as
(R')SAB(rs)ab (de Vaucouleurs et al. 1991), with a Seyfert type 1.5
active nucleus (Osterbrock \& Koski 1976). Optically, the $3'\times
2'$ oval bar and active nucleus are the most prominent features, with
two faint outer spiral arms that show little evidence of star
formation and are visible only in deep optical images (Arp 1977). A
string of H{\sc ii} regions lie along the NW and SE leading edges of
the bar (Schulz 1985). Low-excitation ionised gas associated with
these H{\sc ii} regions was detected in H$\alpha$+N[{\sc ii}]
narrow-band images by P\'erez-Fournon \& Wilson (1990), along with
high-excitation gas in the Narrow Line and Extended Narrow Line
Regions (NLR/ENLR) that is photo-ionised by hard ultraviolet radiation
from the AGN.

Despite the lack of star formation beyond the bar, the faint outer
spiral arms are gas rich and are prominent in neutral hydrogen
images, beginning at each end of the oval bar and winding out to a
radius of 6 arcmins ($\sim$23 kpc) (Davies 1973; Bosma, Ekers \&
Lequeux 1977; Pedlar et al. 1992; Mundell et al. 1999). The oval bar
is also unusually gas rich (Pedlar et al. 1992; Mundell et al. 1999)
and was studied in detail by Mundell \& Shone (1999) who demonstrated
that the gas dynamics of the oval are consistent with those of a
kinematically weak bar. Mundell \& Shone (1999) identified sharp velocity
changes across bright regions close to the leading edges of the bar,
concluding that these regions represent offset shocks, as predicted by
simulations of gas flows in barred potentials (Athanassoula 1992a,b),
which have formed due to the convergence of gas orbital streamlines
associated with the two families of stellar orbits, $x_1$ and $x_2$,
that must be present in the bar.

In this paper we present a study of the distribution and kinematics of
{\em ionised} gas in the arcminute-scale bar of NGC~4151 using
observations of the UV continuum emission at $\lambda$=2905~\AA\
obtained with the UVOT on board XMM-Newton and H$\beta$ emission
obtained with the TAURUS Fabry-Perot interferometer on the Isaac
Newton Telescope (INT). The advantage of TAURUS over long-slit
spectroscopy (e.g. Unger et al. 1987) or narrow band imaging
(P\'erez et al. 1989) is that it simultaneously provides both
complete spatial coverage with a large field of view and
two-dimensional kinematic information (at intermediate velocity
resolution). Fabry-Perot interferometers have been used successfully
to study the distribution and kinematics of ionised gas in a number
 of active galaxies (e.g., NGC5128 - Bland, Taylor \& Atherton 1987;
 NGC1275 - Unger et al. 1990; NGC1068 - Cecil, Bland \& Tully
 1990; Unger et al. 1992) and here we present the first
 Fabry-Perot observations of H$\beta$ emission from NGC~4151. 

The paper is structured as follows: the observations and data
reduction are presented in Section 2, the properties of the
ultraviolet continuum and ionised gas are described in Section 3 and
compared with previous optical observations of this galaxy. In Section
4, we place our results in the context of barred-galaxy gas-dynamic
studies by comparing the distribution and kinematics of the ionised
gas with those previously derived by Mundell \& Shone (1999) for neutral
hydrogen in NGC~4151, concentrating particularly on the behaviour of
the ionised gas in the vicinity of the H{\sc i} shocks that Mundell \&
Shone (1999) identified along the leading edges of the oval bar.

For consistency with previous studies, we use a value of 13.3~Mpc
(H$_0$=75~km~s$^{-1}$~Mpc$^{-1}$) for the distance to
NGC~4151. However, we note that the true distance is likely to be
greater given its membership of the Virgo Cluster (see Mundell {\it et
al.} 1999 for full discussion) and distance-dependent quantities
derived in this paper are therefore lower limits.

\section{Observations \& Data Reduction}

\subsection{Ultraviolet Continuum Imaging}

We obtained archival ultraviolet images taken on 2000 December 22 and
23 (P.I. Griffiths) with the optical monitor on XMM-Newton in the UVW1
($\lambda$$_{c}$$\sim$2905~\AA, $\Delta\lambda$$\sim$620~\AA) filter
(see Mason et al. 2001). The total exposure time was 10.8~ks, the
pixel size is 0\farcs95 and the data were processed using SAS
version~5.4. Measured counts in the final median-combined image were
converted to flux units using a value averaged over different spectral
types as listed in Section 7.1.2.8 (Method 2 by Breeveld) in the SAS
Watchout documentation\footnote{(see
$http://xmm.vilspa.esa.es/external/xmm\_sw\_cal/sas.shtml$)}.

\subsection{TAURUS H$\beta$ observations}

Observations of the H$\beta$ 4861\AA\ line emission from NGC~4151 were
obtained at the f/15 Cassegrain focus of the 2.5-m Isaac Newton
Telescope (INT) using the original version of the TAURUS scanning
Fabry-Perot interferometer instrument (Taylor \& Atherton 1980;
Atherton et al. 1982). The observations used the blue-sensitive
IPCS detector (Boksenberg 1982) which was not well suited to H$\alpha$ (6563\AA) observations.
The 75$\mu$m etalon used had a free
spectral range (FSR) of 15.45\AA\ (953.4 kms$^{-1}$) and was rapidly
scanned in steps of 0.175\AA\ (10.800 kms$^{-1}$) with the data being
summed into a three-dimensional data cube with $350 \times 350$ pixels
(of 0.56 arcsecs$^{2}$ each) in the spatial dimension and 100 pixels
in the spectral dimension. The limited FSR of TAURUS also meant that it would be
problematic to separate H$\alpha$ from [NII]6583\AA\ lines and thus
H$\alpha$ was not observed.

The TAURUS data were reduced using {\sc taucal} (Lewis \& Unger 1991)
running within the {\sc figaro} and {\sc ndprogs} environments of the
{\sc starlink} package. The data were first phase-calibrated
(Atherton et al. 1982) relative to exposures of a neon arc
source taken before the galaxy exposures and then smoothed in order to
improve the signal-to-noise ratio.  Gaussian profiles were then fitted
using {\sc tau\_fits}, a routine within {\sc taucal}, which gave an
estimate of the intensity, central wavelength, width and continuum
level for each pixel with line emission. Finally, maps of the
intensity distribution, velocity field and velocity dispersion were
derived after fitting each of the spectra. Unfortunately, the complex wavelength calibration used for TAURUS data 
cubes does not enable accurate flux calibration or line ratios to be 
achieved. The observational
parameters and final spatial and spectral resolutions for the H$\beta$
data are summarised in Table 1.

\setcounter{table}{0}
\begin{table}
\centering
\caption{TAURUS H$\beta$ observational parameters.}
\begin{tabular}{lc}
\hline

Date of Observation   & 1989 February 25  \\
Integration time(s)   & 3200  \\
Etalon Gap ($\mu$m) & 76.996  \\
Wavelength of order-sorting filter (\AA) & 4883 \\
Bandpass of order-sorting filter (\AA) & 15 \\
Order of interference & 315 \\
Free Spectral Range (\AA) & 15.45 \\
Spectral scale (\AA pixel$^{-1}$) & 0.175 \\
FWHM spectral resolution (\AA) & 0.65 \\
Spatial scale (arcsec pixel$^{-1}$) & 0.56 \\
FWHM spatial resolution (arcsec) & 1.82 \\
Field of view (arcsec$^{2}$) & $196\times 196$ \\

\end{tabular}
\end{table}

\section{Results \& Analysis}

\setcounter{figure}{0}
\begin{figure*}
\setlength{\unitlength}{1mm}
\begin{picture}(-10,100)
\put(0,0){\includegraphics{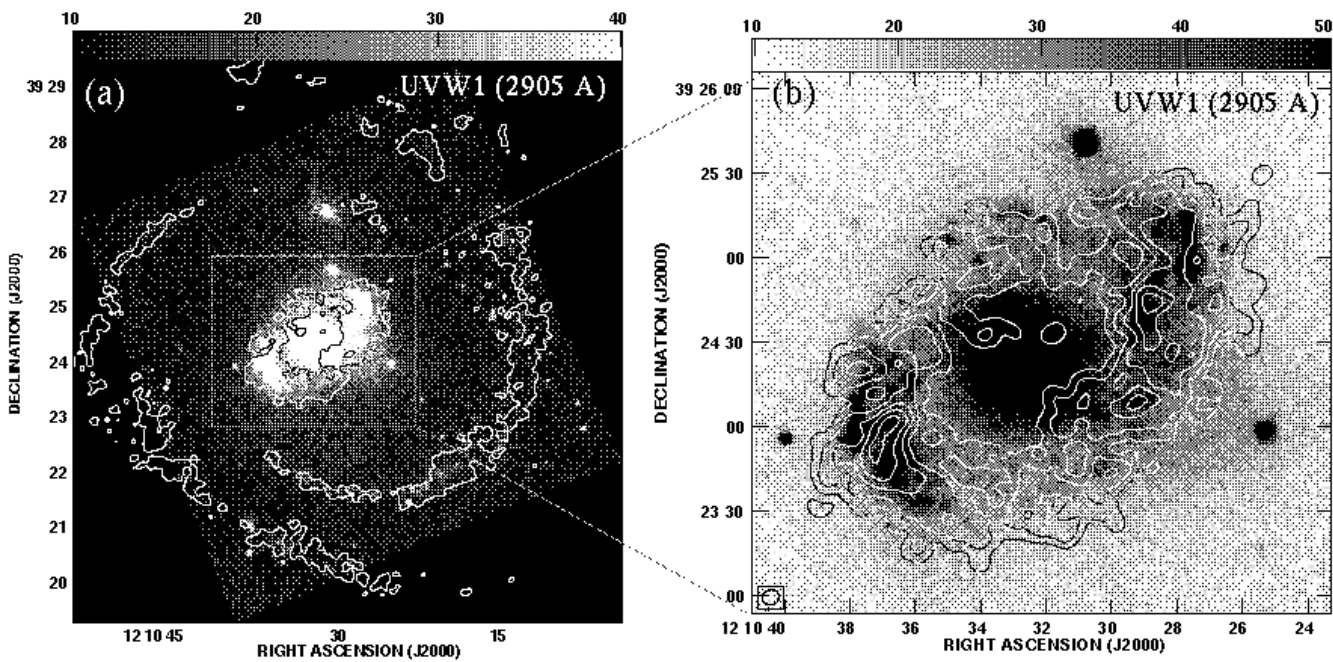}}
\end{picture}
\caption{XMM-Newton ultraviolet image of NGC~4151 in the UVW1 filter 
with H{\sc i} contours from Mundell et al. (1999) overlaid: (a) whole galaxy (b) oval bar
region. The flux density scale is shown at the top of each
image in units of 10${\rm ^{-16}~erg~s^{-1}~cm^{-2}~\AA^{-1}}$; H{\sc i}
contour levels correspond to column densities (a)
N$_H$=~0.4$\times$10$^{21}$~cm$^{-2}$ and (b) N$_H$= (0.3, 0.8, 1.8,
2.8, 3.6)~$\times~$10$^{21}$~cm$^{-2}$ and the beam-size is 6$\arcsec$$\times$5~$\arcsec$.  
}
\label{uvhi}
\end{figure*}

The oval bar and AGN are bright at ultraviolet wavelengths; fainter UV
emission is also detected coincident with the outer spiral arms,
particularly towards the bluer south/south-west arm.  Within the oval
bar, spatially-resolved H$\beta$ emission is detected from three main
regions: faint emission from H{\sc ii} regions that lie along the NW
and SE leading edges of the bar and stronger emission associated with
NLR and ENLR, elongated along P.A. $\sim$48$^{\circ}$, consistent with
structures identified in earlier H$\alpha$+N[{\sc ii}] narrow-band
images (e.g. P\'erez-Fournon \& Wilson 1990).

\subsection{Ultraviolet Continuum Emission}

Figure \ref{uvhi} shows XMM-Newton ultraviolet continuum images in the
UVW1 (2905 \AA) filter of NGC~4151, with contours of H{\sc i} emission
overlaid to illustrate the structure of the galaxy. As can be seen in
Figure \ref{uvhi}(a), the oval bar and AGN are prominent but faint
emission from the outer spiral arms is also detected. Ultraviolet
continuum emission from the spiral arm that begins at the SE end of
the bar can be followed to more than half way along its length, while
faint patches associated with arm that begins at the NW end of the bar
can be seen particularly towards its southern-most tip; the
ultraviolet continuum is well traced by the H{\sc i} structures,
suggesting that star formation is occuring at a low level in the
regions of highest H{\sc i} column density and that the neutral gas is
porous.

In the bar, the bright regions associated with the H{\sc ii} regions
along the NW and SE leading edges are prominent (Figure \ref{uvhi}(b))
and broadly coincide with the regions of highest H{\sc i} column
density. Fainter, diffuse ultraviolet emission is present across the
entire bar and despite its patchy nature, the UV emission clearly
de-lineates two curved arms winding clockwise from the leading edges of
the bar in towards the nucleus, in a remarkably similar way to that
observed in H{\sc i} (Mundell \& Shone 1999; Mundell et al. 1999). The
total luminosities, L$_{UV}$ (uncorrected for extinction), measured in
two elliptical apertures (semi-major axis 27\farcs6, ellipticity 0.73)
centred on the NW and SE H{\sc ii} regions are L$_{UV}$~$\sim$~1.5 and
1.2~$\times$~10${\rm ^{38}~erg~s^{-1}~\AA^{-1}}$ respectively.
Archival ultraviolet images of NGC~4151 taken with the Ultraviolet
Imaging Telescope (UIT) are not sensitive to the faint UV emission
in the spiral arms but do confirm the structure and blue colour of the bar
HII regions as seen in our XMM-Newton image (e.g. Fanelli et
al. 1997; Kuchinski et al. 2000

\subsection{H$\beta$ Emission}

\begin{figure*}
\setlength{\unitlength}{1mm}
\begin{picture}(-10,160)

\put(0,0){\includegraphics{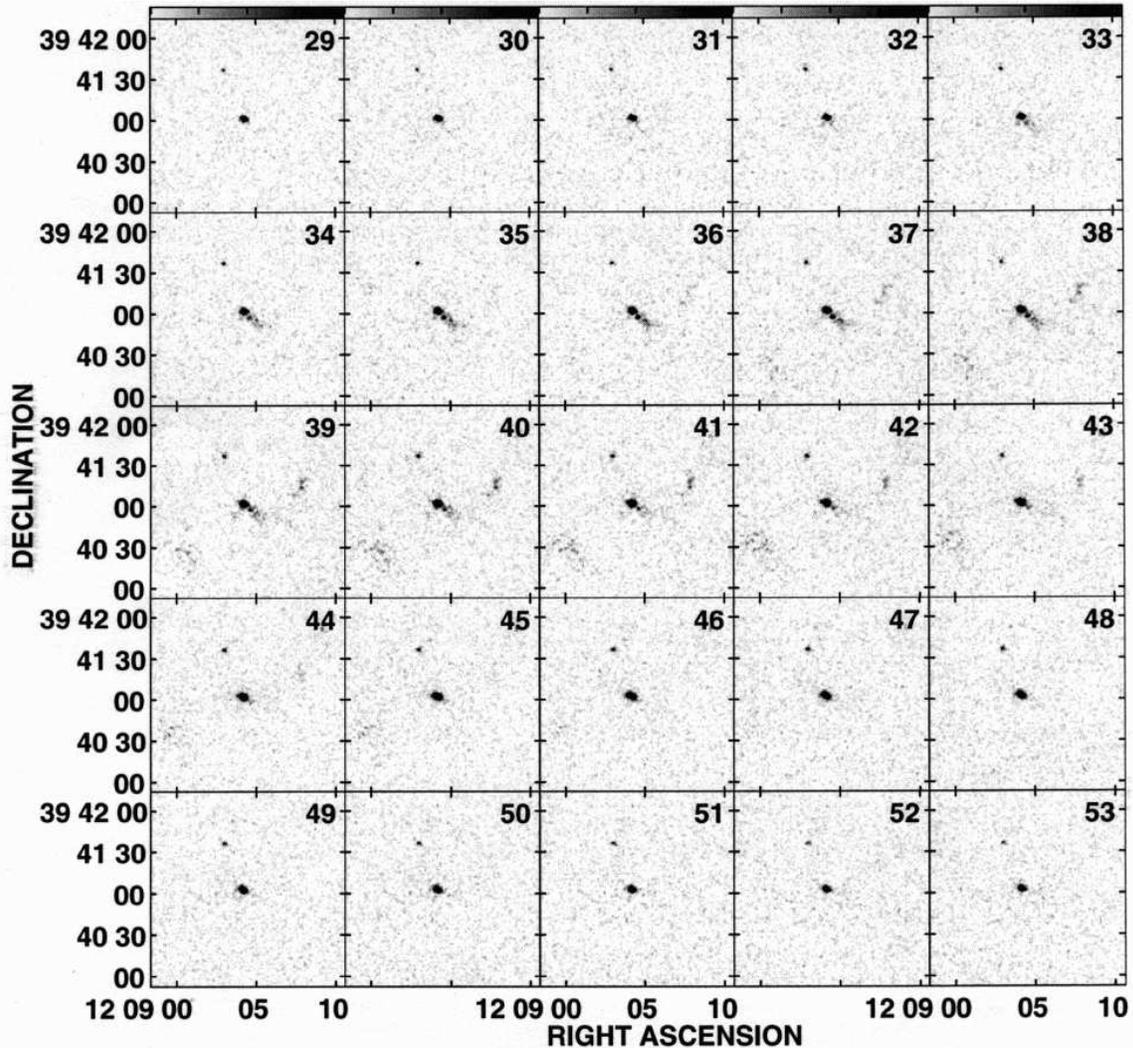}}
\end{picture}
\caption{H$\beta$ velocity channel maps of the H{\sc ii} regions in NGC~4151. The range
 of
values are 946 kms$^{-1}$ to 1065 kms$^{-1}$ with channel 41 corresponding to 998 kms$^{-1}$.
 North is up and East to the left.}
\label{hbchans}
\end{figure*}

Figure \ref{hbchans} shows channel maps from the TAURUS H$\beta$ cube
covering the 3$'$~$\times$~2$'$ oval bar; H$\beta$ emission is
detected from the AGN in all velocity channels, consistent with the
large line-width of the nuclear emission (Robinson et al. 1994). H$\beta$ emission associated
with the NLR and ENLR is detected along P.A.~$\sim$48$^{\circ}$ over
the velocity range 854$-$1123~km~s$^{-1}$ out to a radius of $\sim$25~$\arcsec$
from the nucleus. Finally, faint clumpy H$\beta$ emission can be seen
NW and SE of the nucleus over the velocity range 945$-$1082 km~s$^{-1}$,
coincident with the bar H{\sc ii} regions discussed in the previous
section.
No correction for stellar absorption in the H$\beta$ emission line profiles
has been applied since the narrow free spectral range (15\AA) of TAURUS
means that the data are insensitive to any broad absorption line
components similar to that detected in NGC~1068 (FWZI~$\sim$80\AA) by
Bland-Hawthorn, Sokolowski \& Cecil (1991). To verify this, we inspected
the NGC~4151 H$\beta$ line profiles in long-slit spectra published by Robinson
et al. (1994) in which there is no comparable absorption component. We
therefore conclude that any H$\beta$ stellar absorption will not have a
significant effect on our H$\beta$ image and derived kinematics.

\begin{figure*}
\setlength{\unitlength}{1mm}
\begin{picture}(10,180)
\put(0,0){\includegraphics{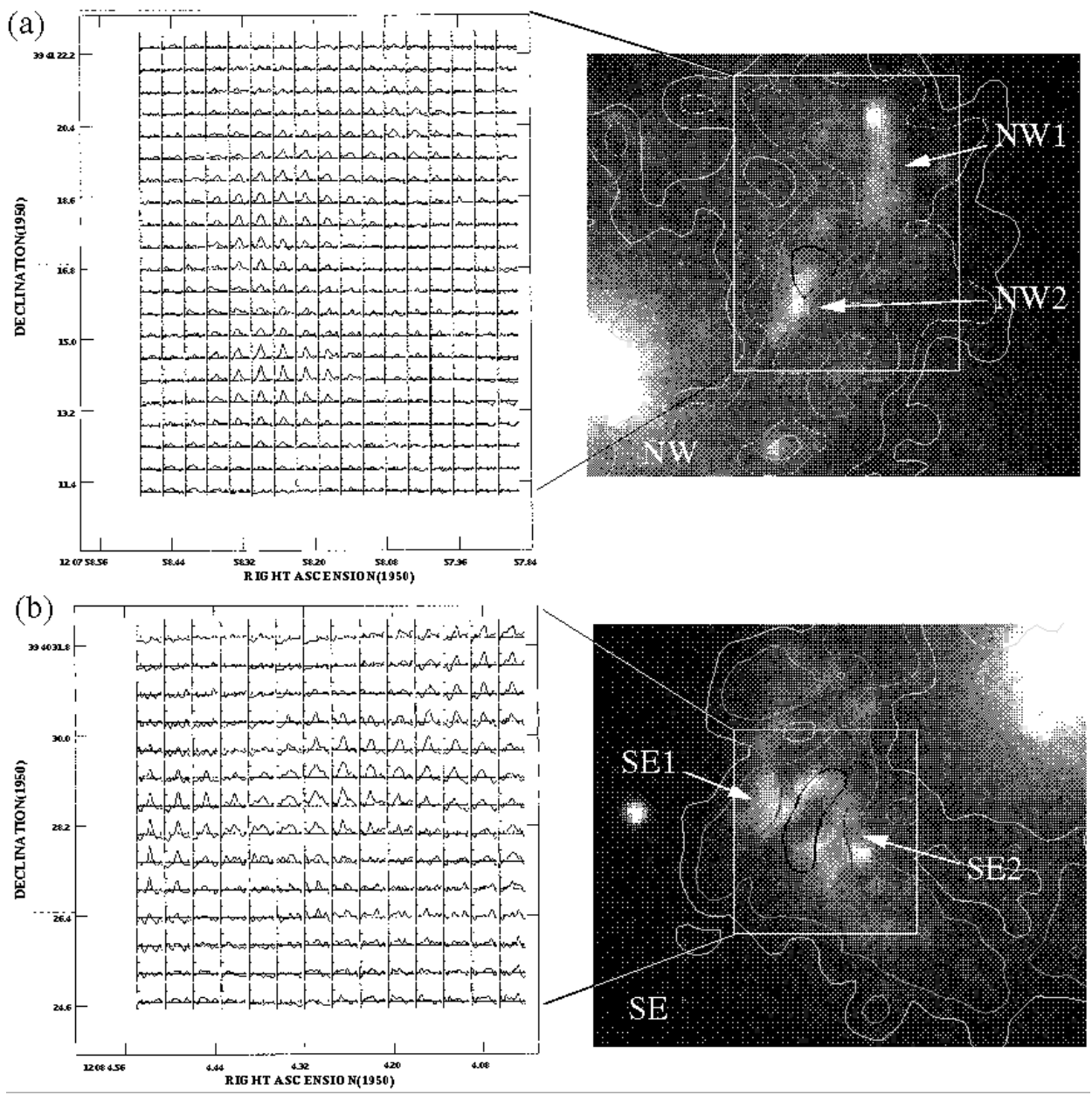}}       
\end{picture}
\caption{{\bf Left:} H$\beta$ spectra of the H{\sc ii} regions along the {\bf (a)}
north-west and {\bf (b)} south-east edges of the bar. Each spectrum
covers a velocity range of 900~$-$~1100 kms$^{-1}$. {\bf Right:}
the UV image of
each region with H{\sc i} intensity contours overlaid
to illustrate the structure of the regions in which H$\beta$ emission
is detected. Individual bright segments in each region are labelled
NW1, NW2, SE1 and SE2 (see text for discussion).
}
\label{shockspec}
\end{figure*}

\subsubsection{The H{\sc ii} Regions}

Figure \ref{shockspec} shows representative spectra of the H$\beta$
emission in the NW and SE regions of the bar. The ultraviolet
continuum and H{\sc i} is also shown for comparison. Although the UV
continuum structure is broadly consistent with two arc-like arms along
the leading-edges of the bar (Figure \ref{uvhi}(b)), the detailed
structure is more complicated. The brightest structures in the NW
region consist of two dislocated arcs, labelled NE1 and NE2 in Figure
\ref{shockspec}, while the bright structure in the SE region resembles
an $\Omega$-like ridge (see also Schulz 1985), each side of which we
label SE1 and SE2 (Figure \ref{shockspec}). 

\begin{figure}
\setlength{\unitlength}{1mm}
\begin{picture}(10,210)

\put(0,0){\includegraphics{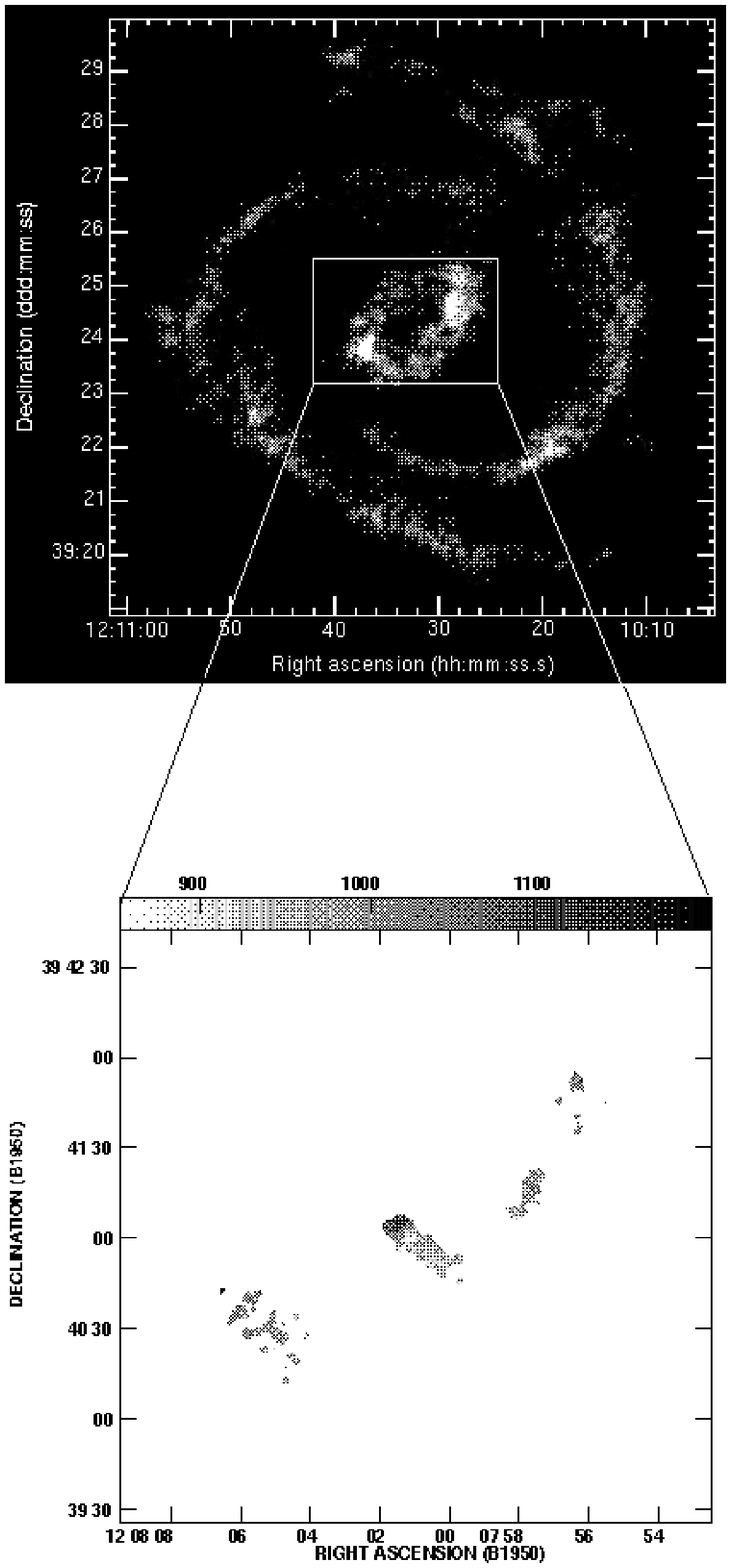}}
\end{picture}
\caption{(a) The H{\sc i} intensity distribution of NGC~4151 from
zeroth moment analysis (Mundell et al. 1999). (b) The velocity distribution of H$\beta$ emission line gas in the nucleus of NGC~4151 derived from TAURUS observations. The range of values are 854$-$1197 km~s$^{-1}$.}
\label{momnts}
\end{figure}

\begin{figure}
\setlength{\unitlength}{1mm}
\begin{picture}(10,110)

\put(0,0){\includegraphics{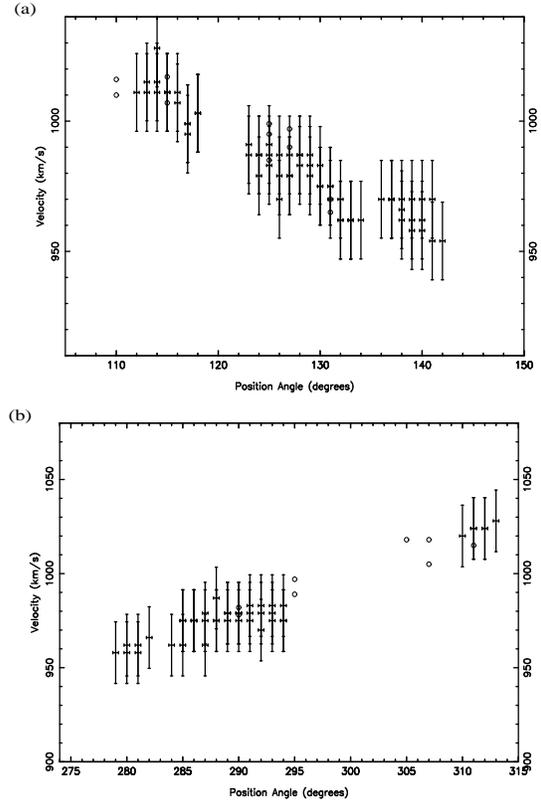}}
\end{picture}
\caption{Optical velocities of the {\bf (a)} SE and {\bf (b)} NW H{\sc ii} regions
 in NGC~4151. The velocities were measured through
azimuthal angles from the nucleus.  TAURUS H$\beta$ velocities are denoted by
crosses and agree well with velocity measurements from Schulz (1985) (open circles).}
\label{optaz}
\end{figure}

As can be seen in the H$\beta$ moment map in Figure \ref{momnts}(b),
which was derived by fitting Gaussian profiles to each spectrum in the
cube, ionised gas is detected in each of these four bright continuum
segments, NW1, NW2, SE1, SE2. The H$\beta$ kinematics are
also consistent with those derived from earlier optical long-slit
spectroscopy; Figure \ref{optaz} shows a comparison between velocities
derived by Schulz (1985) from long-slit spectra of the brightest knots
in the NW and SE H{\sc ii} regions and those derived from the TAURUS
H$\beta$ cube. Radial velocities are plotted as a function of
azimuthal angle, with the AGN at the origin.

\begin{figure}
\setlength{\unitlength}{1mm}
\begin{picture}(10,100)

\put(0,0){\includegraphics{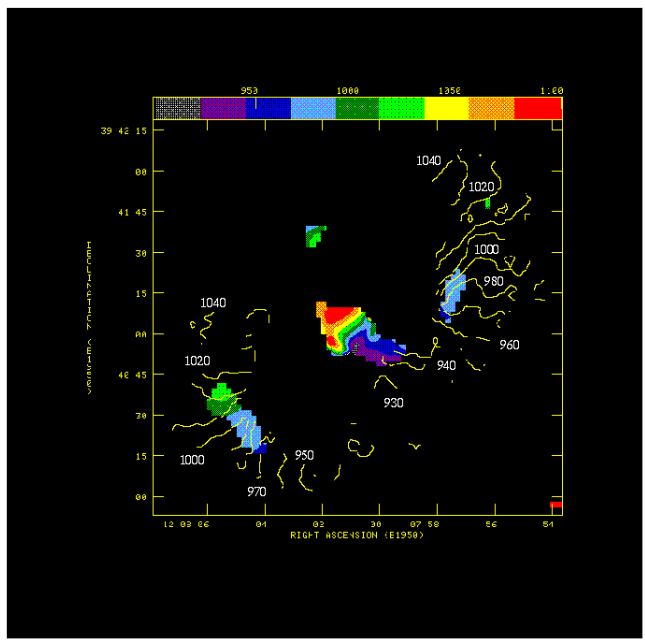}}
\end{picture}
\caption{Comparison of H$\beta$ (colour) and H{\sc i} velocities (contours) of the 
bar region in NGC~4151. The velocities are in kms$^{-1}$.}
\label{hihbvel}
\end{figure}

Figure \ref{hihbvel} shows H{\sc i} iso-velocity contours from Mundell et
al. (1999) overlaid on the H$\beta$ velocity map smoothed to the same
angular resolution (6$"$~$\times$~5$"$); it is clear that the ionised gas
velocities in the bar H{\sc ii} regions closely resemble the cold gas
kinematics which are dominated by rotation within the galactic potential
(see also Figure 5 of Mundell et al. 1999). The larger concentration of H{\sc i} in
the bar compared to ionised gas helps to provide a context in which to
understand the ionised gas features. Firstly, the arc of neutral gas
along the NW edge of the bar appears to join smoothly to the ionised
gas at the SE tip of the ENLR, with ionised and neutral gas velocities
and line-widths matching closely (Asif et al. 1997); this continuity
suggests a spatial and dynamical link between the neutral and ionised
gas and transportation of gas towards the AGN (Mundell \& Shone
1999). A similar correspondence between the NW H{\sc i} and ultraviolet
arc can be seen in Figure \ref{uvhi}(b).

\subsubsection{The Extended Narrow Line Region}

The ENLR in NGC~4151 has already been well-studied using both
narrow-band imaging (e.g. P\'erez-Fournon \& Wilson 1990) and
long-slit spectroscopy (e.g. Unger et al. 1987; Asif et al. 1997). Our
2-D TAURUS observations are consistent with these previous studies.
The velocity profile along the axis of the ENLR was extracted by
interpolating the velocity field across the nucleus; a systemic
velocity of 1000$\pm$2 kms$^{-1}$ was derived, consistent with the
value of 998 kms$^{-1}$ derived by others (Pedlar et al. 1992;
Mundell et al. 1999; Robinson et al. 1994; Asif et al.
1997; Winge et al. 1999).

The H$\beta$ emission lines in the ENLR (i.e. at distances greater
than 5$\arcsec$ from the nucleus) are narrow (FWHM $<$ 25~km~s$^{-1}$) and
consistent with galactic rotation (see also Pedlar et al.  1992;
Robinson et al. 1994; Vila-Vilar\'o et al.  1995; Asif
et al. 1997; Winge et al. 1999). Indeed, as shown in
Figure 4(b) there is a clear distinction between the velocity
structure of the NLR, which is dominated by outflow (e.g. Hutchings et al.
1999; Crenshaw et al. 2000) and the ENLR. However, our TAURUS data show some
evidence for additional velocity structure within the SW ENLR, with
measured deviations from circular velocity of up to 20 kms$^{-1}$,
which could correspond to much higher intrinsic velocities when
corrected for galactic inclination.
Evidence for a relatively wide angle cone (opening angle of $\sim$120~degrees) partially intercepting the 
galactic disk (to produce the ENLR) has been discussed extensively by Pedlar et al.
(1992), Robinson et al. (1994) and Vila-Vilar\'o et al. (1995).

\section{Discussion}

The location of ionised gas (H$\alpha$), molecular gas (CO) and dust
in barred galaxies is often observed to be offset along the leading
edge of the host stellar bar (e.g. Martin \& Friedli 1997; Ondrechen
1985; Handa et al. 1990; Regan \& Vogel 1995; Reynaud \& Downes
1998). Sheth et al. (2002) find a further offset between the
distribution of molecular gas and H{\sc ii} regions as traced by
H$\alpha$ emission, with the H$\alpha$ emission generally offset
towards the leading edge of the CO emission, and the largest offsets
found in the strongest bars. These offsets are interpreted broadly as
evidence for stellar clusters forming in compressed pre-shock gas or
molecular cloud complexes upstream from the shock; these stellar
clusters then travel through the shock ballistically on elliptical
orbits and ionise the post-shock gas. This is in contrast to the
dissipative neutral or molecular gas in the bar that is focussed and
redirected inwards when it meets the shock. In NGC~4151, although no
published detection of molecular gas in the bar exists, neutral
hydrogen is sharply focussed and streams inwards along the
leading-edge shocks (Mundell \& Shone 1999). Here we examine the
distribution and kinematics of the ionised gas with respect to that of
the neutral gas and discuss whether obscuration, kinematics or a
combination of both might explain the observed features.
In these observations, the optical astrometry was tied to the position of the continuum nucleus 
of NGC~4151 and was accurate to 1$\arcsec$ at least. This is 
adequate for comparison with the H{\sc i} emission which has an angular 
resolution of $\sim$6~$\arcsec$.

\subsection{Origin of Ionised Gas in the Bar Shocks}

Overall, the UV light distribution in the bar of NGC~4151 is patchy;
the whole bar is filled with diffuse UV light, with the bright H{\sc ii}
regions along the leading edges of the bar standing out clearly. Spurs
and knots of UV emission are also evident close to the minor axis of
the bar, particularly $\sim$45~$\arcsec$ north of the galaxy nucleus. As
shown in Figures \ref{uvhi}(b) and \ref{shockspec}, the H{\sc ii}
regions in the bar are concentrated around the H{\sc i} shocks; in the
SE, the small arc of highest H{\sc i} column density is coincident
with the dark region in the $\Omega$-shaped ridge, so that the bright
UV/H$\beta$-emitting regions SE1 and SE2 straddle the H{\sc i}
ridge. In the NW, the two bright UV ridges, NW1 and NW2 are similarly
separated by an apparent dust lane that lies along the H{\sc i} shock;
specifically, the NW1 UV ridge is located azimuthally downstream of the
H{\sc i} shock ridge. However, the NW2 ridge appears to line up
closely with that of the H{\sc i}, although the actual emission peaks
do not coincide identically.

The patchy nature of the UV light may in part be explained by
extinction.  Assuming the H{\sc i} forms a screen in front of the UV
emission, the maximum amount of obscuration due to the H{\sc i}, which
has a peak column density in the shocks N$_{\rm
H}$~=~3.6~$\times$~10$^{21}$~cm$^{-2}$, corresponds to 4.9 magnitudes
of extinction at $\lambda$=2905\AA\ (assuming a Galactic extinction
law; Staveley-Smith \& Davies 1987). However, mixing of the gas and
stars is more likely making this value an upper limit. Nevertheless,
obscuration in the H{\sc i} shocks, which are also likely to be
dust-rich, may partially explain the patchy UV emission and the
appearance that the ionised-gas/UV-bright regions are
straddling/bracketing the H{\sc i} shocks, particularly in the SE.
The likely effect of dust on the H$\beta$ profiles is difficult to tell due
to the lack of any H$\alpha$ data.
The profiles shown by Robinson et al. (1994) do not show significant
differences between the two lines and we do not expect the 
effect to change the kinematics dramatically. However, these profiles are not from the
shocked regions and also have relatively poor velocity resolution (90 kms$^{-1}$ to 420 kms$^{-1}$).
A better test for this effect would be to use high velocity resolution
(6 kms$^{-1}$) echelle spectroscopy of both H$\alpha$ and H$\beta$ with high signal-to-noise. 

\begin{figure}
\setlength{\unitlength}{1mm}
\begin{picture}(10,90)

\put(0,0){\includegraphics{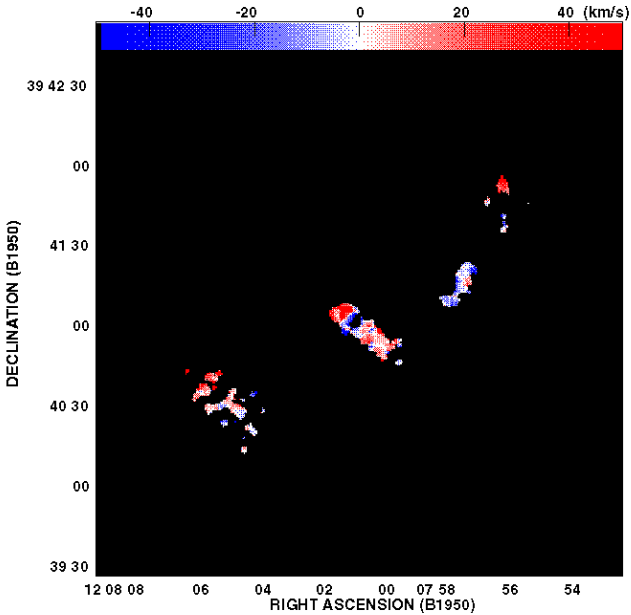}}
\end{picture}
\caption{H$\beta$ residual velocity field; colour scale units are km~s$^{-1}$.}
\label{hbresidual}
\end{figure}

Alternatively, dynamical effects may explain the distribution of
ionised gas. Mundell \& Shone (1999) showed that the neutral hydrogen
is distributed throughout the oval bar, with strong streaming motions
concentrated along the highest column density H{\sc i} ridges, evident
in the residual velocity field derived by removing the circular
rotation component, and confirming their identification as 
leading-edge shocks. N-body/SPH simulations by Sheth et al. (2002)
showed that compression of gas clouds travelling into the shock region
can trigger star formation upstream of the shock; these stellar
clusters then travel ballistically through the shock region, unlike
the gas which is channeled inwards along the shock towards the
nucleus, to ionise ambient gas beyond the shock. Taking a lifetime of
an H{\sc ii} region to be 3 Myr (Spitzer 1978), Sheth et al. (2002)
assume a lifetime of 10 Myr for the H$\alpha$ emission which
represents a spread in H{\sc ii} region ages given that individual
H{\sc ii} are not spatially resolved. 

For NGC~4151, we use the bar pattern speed 24.5~km~s$^{-1}$~kpc$^{-1}$
and galaxy rotation speed derived by Mundell et al. (1999) to derive
the speed at which newly-formed stars will enter the bar shocks, at
$\sim$3.9~kpc from the galaxy centre. For a lifetime of 10~Myr,
consistent with the UV luminosity, the maximum distance travelled is
1.2~kpc or 18$\arcsec$. This is distance is consistent with the total
extent of the ionised gas across both shock regions. To further test
this dynamical scenario we compared the 2-D ionised gas velocities
measured in our TAURUS cube with those measured by Mundell \& Shone
(1999) for the H{\sc i}; as shown in Figure \ref{hihbvel}, the H{\sc i}
and H$\beta$ velocities are consistent and are therefore dominated by
the rotation of the galaxy. In order to examine departures from
circular rotation we derived an ionised gas residual velocity field
using the same method that Mundell \& Shone (1999) used to derive the
H{\sc i} residual velocity field (their Figure 2).  We used the H{\sc
i} rotation curve (Mundell et al. 1999) to derive a 2-D model velocity
field which was then subtracted from the ionised gas velocity field to
leave the residual velocity field shown in Figure \ref{hbresidual}. We
have applied the same colour table as that used by Mundell \& Shone
(1999) to allow a direct comparison of ionised and neutral gas
velocity residuals. Although spatial coverage of the ionised gas is
limited compared with that of the neutral gas, it is clear that the
two residual fields are consistent; comparison of the ionised and
neutral gas residual velocity fields suggest that NW1 is (red)
post-shock ionised gas, NW2 is (bluer) shocked ionised gas streaming
towards the nucleus similar to the H{\sc i} shock; SE2 is (blue)
pre-shock gas entering the shock and SE1 is (red) shocked gas similar
to that seen in H{\sc i}. 

\begin{figure}
\setlength{\unitlength}{1mm}
\begin{picture}(10,90)

\put(0,0){\includegraphics{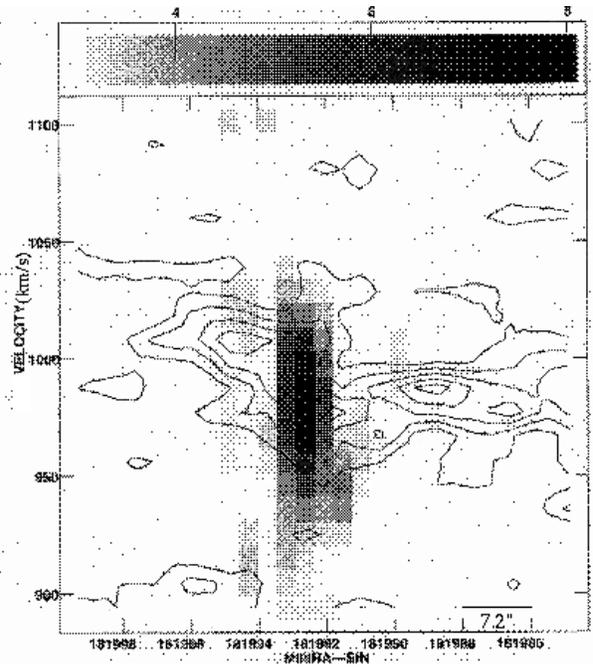}}
\end{picture}
\caption{H$\beta$ (grey-scale) and H{\sc i} (contour)
position-velocity diagrams extracted using a 6$\arcsec$-wide slit centred on the H$\beta$ peak in NW2
and perpendicular to the direction of the shock ridge (see Mundell \&
Shone 1999). }
\label{nw2hihb}
\end{figure}

Although the majority of the bright UV emission and associated ionised
gas is consistent with pre- and post-shock gas, the NW2 region may
represent the unusual case of ionised gas {\em in} a bar shock. NW2
coincides with the region of highest N$_{\rm H}$ and strongest H{\sc
i} streaming (Mundell \& Shone 1999). 
The much larger concentration of H{\sc i} in the bar compared to
ionised gas allows greater opportunity to look for the sudden jumps in
velocity characteristic of gaseous shocks. Nevertheless, we used the
same method as Mundell \& Shone (1999) to extract the H$\beta$
emission in a 6$\arcsec$-wide slit centred on the H$\beta$ peak in NW2
and perpendicular to the direction of the shock ridge in order to
search for the predicted maximum velocity jump in the ionised gas.
Figure \ref{nw2hihb} shows the resultant position-velocity diagrams of
H$\beta$ (gray-scale) and H{\sc i} (contour) superimposed; this figure
shows that the steep velocity gradient identified in H{\sc i} data of
Mundell \& Shone (1999) is also visible in the H$\beta$ data,
further confirming that NW2 represents ionised gas in the shock itself.

We therefore conclude that the distribution of bright UV emission is
broadly consistent with the formation of star clusters upstream
travelling ballistically through the gaseous shock to ionise gas
downstream along the leading edge of bar. In addition, we suggest that
ionised gas within the shock itself is detected for the first time and
is streaming to smaller radius orbits in the same manner as the
neutral gas. Indeed, Reynaud \& Downes (1998) find that massive stars
cannot form in bar shocks in which the velocity shear is high, as is
the case in the H{\sc i} shocks in NGC~4151 (Mundell \& Shone 1999);
this further supports our interpretation that although we detect
ionised gas in the shock regions, it cannot be ionised by stars
formed in situ but instead is ionised by stars formed upstream as
they pass ballistically through the shock region.  The observed
kinematics of the brightest H{\sc ii} regions as measured by Schulz
(1985), which are consistent with elliptical stellar orbits, are then
easily explained.

\section{Conclusions}

The TAURUS Fabry-Perot interferometer was used to simultaneously map the
 distribution and kinematics of H$\beta$ emission in the Seyfert 1 galaxy NGC4151. The results were compared with that previously
 determined in neutral hydrogen by Mundell \& Shone (1999). We detect for the first time, ionised gas within the shock itself which is streaming to smaller radii in the same manner as the neutral gas.

We suggest that the
 distribution of bright, patchy UV emission close to the H{\sc i}
 shocks is, on the whole, consistent with ionisation by star clusters that
 have formed in compressed pre-shock gas as
they pass ballistically through the shock region. The kinematics are consistent with elliptical stellar orbits.
The H$\beta$ ENLR velocity profile is consistent with previous studies.

%

\section{Acknowledgements}

The authors are grateful to Steve Unger for the TAURUS observations and 
for his support of this project in its early stage.
We thank Dave Shone and Johan Knapen for useful discussions and Jim Lewis
for advice on TAURUS data reduction.
CGM acknowledges financial support from the Royal Society. We thank the anonymous referee for constructive comments that
helped to improve the paper.
The INT is operated on the island of La Palma by the Isaac Newton Group in the Spanish
Observatorio del Roque de los Muchachos of the Instituto de Astrofisica de Canarias.

\end{document}